# THERE'S NO PLACE LIKE HOME (IN OUR OWN SOLAR SYSTEM):
## Searching for ET Near White Dwarfs

**JOHN GERTZ** Zorro Productions, 2249 Fifth Street, Berkeley, CA, 94710, USA

**Email** jgertz@zorro.com

The preponderance of white dwarfs in the Milky Way were formed from the remnants of stars of the same or somewhat higher mass as the Sun, i.e., from G-stars. We know that life can exist around G-stars. Any technologically advanced civilization residing within the habitable zone of a G-star will face grave peril when its star transitions from the main sequence and successively enters sub-giant, red giant, planetary nebula, and white dwarf stages. In fact, if the civilization takes no action it will face certain extinction. The two alternatives to passive extinction are (a) migrate away from the parent star in order to colonize another star system, or (b) find a viable solution within one's own solar system. It is argued in this paper that migration of an entire biological population or even a small part of a population is virtually impossible, but in any event, far more difficult than remaining in one's home solar system where the problem of continued survival can best be solved. This leads to the conclusion that sub-giants, red giants, planetary nebula, and white dwarfs are the best possible candidate targets for SETI observations. Search strategies are suggested.

## 1  INTRODUCTION: WITH SO MANY HABITABLE PLANETS, WHERE IS ET?

Some have argued that great advantages would accrue to extraterrestrial (ET) civilizations that utilized physical probes to gather information, send information and create an information sharing network, obviating reliance on electromagnetic (EM) beacons transmitted from ET's home planet [1-9]. However, this begs the question of why ET would send probes but would not itself visit or, more ominously, colonize Earth. For the purposes of this paper, the ET in question are biological beings rather than AI robots.

Enrico Fermi's famous 1950 question, "so, where is everyone?" pre-dated the first modern SETI experiment by ten years. It was premised on the dual observation that (a) since interstellar travel is not forbidden by the laws of physics, and (b) that because ET could have had billions of years in which to arrive, ET should even now be found within our own Solar System. About 25 years later, the term, "Fermi's Paradox" came to be more broadly applied not just to this question of the apparent absence of ET spaceships, but also to the failure to detect ET's EM beacons.

Sagan proposed that ET would colonize, but only in such a limited manner that most of the Galaxy would be left untouched [10]. Jones, on the other hand, was led to conclude that colonization should permeate the Galaxy and that therefore "unless we discover that interstellar travel is impractical, I conclude that we are probably alone in the Galaxy" [11]. If an urge to explore or expand or if local resource depletion is insufficient to spur colonization, then the imminent death of one's star should act as a prime motivator. Any ET that had evolved around a G-star would face the inevitable mortality of its home planet once its star leaves the main sequence and becomes a red giant (RG). However, if it can be shown that despite this bleak outlook (a) successful interstellar colonization is nonetheless unachievable and/or (b) in any event, it is eminently possible to survive and thrive during the RG and white dwarf (WD) phases *in situ*, then SETI efforts might best concentrate on RGs and WDs. For if they are able to survive the departure of their star from the main sequence, then it is around RGs and WDs that the galaxy's most advanced civilizations would reside, and it is there where vast engineering projects, necessitated by their star's transition, may be detectable, whether or not ET is currently transmitting beacons to other civilizations. It is the purpose of this paper to demonstrate that interstellar travel may indeed be so impractical that for all intents and purposes, it may never be successfully undertaken. This, in turn, will lead to suggestions for major revisions in prevalent SETI search strategies.

## 2  RECAPPING THE STANDARD ARGUMENTS AGAINST COLONIZATION

When constructing an interstellar space vehicle, there is a fundamental tradeoff between the energy requirements to achieve very fast velocities and the duration of flight. A flight at relativistic speeds might be preposterously expensive in terms of

---

This paper was presented at the Foundations of Interstellar Studies Workshop 2 at Charfield, UK.

energy, while flights at the relatively slow velocities we can imagine using understood, though futuristic technologies, would take preposterously long periods of time for biological beings such as humans (i.e., thousands or tens of thousands of years) to reach even the nearest stars [12].

## 2.1 World Ships May Be Too Fuel Intensive

World ships have been imagined that would transport large populations of settlers to new homes across interstellar distances [for a review, 13]. The general size imagined in the literature is in the hundreds of square km of internal living space, holding thousands to hundreds of thousands of human space voyagers. It seems reasonable to imagine that an AI probe might range in size from a postage stamp to a school bus, whereas a colonizing world ship, housing multiple generations of inhabitants, would reasonably be many orders of magnitude larger, perhaps the size of a city plus surrounding countryside for agriculture, plus very ample space for an ark full of life forms destined for introduction on the target planet.

To give a sense of the size of what is proposed, a modern aircraft carrier is populated by approximately 6000 crew members living in extremely tight quarters. A world ship for just 6000 inhabitants, i.e., a very tiny fraction of the Earth's population, would need to be vastly larger for it must allow for much larger and more comfortable quarters for a journey lasting millennia; shielding; space for vast quantities of supplies that unlike in the case of an aircraft carrier cannot be readily replaced at ports of call; raw materials for manufacturing and repairs; large areas for agriculture and more yet for water; and space for the menagerie of plants and animals that would be transplanted on the destination planet. Shielding against cosmic rays would itself require that the ship be cocooned in the equivalent of three feet of concrete. The mass of propellant, depending upon the method and efficiency of propulsion and the desired cruising velocity, will in all cases be much larger than the mass of the ship itself.

Forward reviewed eight more or less viable methods for accelerating a world ship to reasonable interstellar velocities, including by means of hydrogen bombs being expelled and exploded out the back end at 3 second intervals, light sails, matter/anti-matter engines, and pellet streams [14]. All of these methods imagine technology well in advance of current capabilities, and some of these methods, though theoretically possible, may never be feasible in practice. For example, we have no idea how to make and contain massive amounts of antimatter for a matter/antimatter engine. Similarly, the rapid explosion of hydrogen bombs begs the question of how the world ship occupants would survive the kinetic impact of the explosions and the intense gamma rays they would produce. The crucial point is that the greater the desired cruising velocity the greater the amount of propellant that must be included, with that amount increasing exponentially for large relativistic velocities. Even were world ship acceleration to relativistic velocities achievable using extremely advanced engineering, it might nevertheless court disaster when the world ship inevitably collides with interstellar dust or micrometeorites.

## 2.2 World Ships May Be Too Time Intensive

There may simply be no feasible means to accelerate a city-sized world ship to relativistic speeds. The Daedelus study concluded that a realistic, though still far futuristic engine, might accelerate a spacecraft to a cruising velocity of .14 c such that it could arrive at one of Sol's closest neighbor, Barnard's Star (5.9 ly) in about 50 years. However, this was conceived as a flyby. A mission that entailed deceleration would take about 100 years. Even this presupposes great advances beyond current technology, given that Earth's fastest interstellar spacecraft to date, Voyager I, would take more than 105,000 years to travel the same distance. However, Daedelus, like Voyager I was conceptualizing a space probe not a world ship. Voyager I weighs just 733 kg. This is many orders of magnitude smaller than a human bearing world ship [15,16]. In human terms, a world ship traveling at the velocity of Voyager I would require about 4,200 generations to reach Barnard's Star. All but the very first and the very last of these generations would be born, live their entire lives, and die en route. It staggers the imagination to think of the life support systems that would be required to work flawlessly over such deep time. One hundred percent recycling efficiency would be required, since raw materials could not be gathered in interstellar space, even from comets in the Oort cloud, except at an enormous energetic price paid in order to decelerate to a stop, relative to the comet, and then reaccelerate to interstellar cruising speeds. Inhabitants would have to practice rigorous population controls. All repair parts would need to be fabricated from brought materials. Frozen sperm, eggs, or zygotes would have to be included in the prelaunch inventory to avoid inbreeding. Although ET might not be a social species, for humans or any similar social animal, tribalism, conflict, loss of perception about the mission, degradation of necessary skills, and the philosophical, political, and religious changes over deep time might preclude any possibility of success. Even imagining that a world ship could achieve a relativistic velocity, inhabitants might still be forced to spend tens or hundreds of generations within the world ship at its destination while a taget planet is terraformed (considered more fully below).

## 3 BEYOND THE ENERGY/TIME TRADEOFF: IS INTERSTELLAR COLONIZATION FANTASTIC OR JUST FANTASTICAL?

Classical arguments against interstellar colonization based upon the tradeoff between energy and time are strong, but nonetheless, not fatal to the proposition. However, there are other potential failure points:

## 3.1 Perhaps ET is Advanced, But Not So Advanced

Given that ET might have become technological at any time within at least the last 5 billion years, the chances of encountering any civilization that is as newly able to communicate across interstellar space as ourselves is statistically a virtual impossibility. Therefore, in accordance with Arthur C. Clark's oft-quoted maxim, ET's science will seem like magic to us. However, ET's "magic" will have its limits in that ET will be constrained by the same laws of nature as ourselves. It cannot travel faster than light any more than we can; and gravity for ET will be the same as it is for us. ET's technology may seem magical at first, but might rely at its core on some of the same features as ours. In fact, much of our most useful technology today is based upon such pre-historic inventions as thread and cloth, agriculture, fire, the wheel, domesticated animals, buildings, metallurgy, tools such as knives and hammers, ceramics, glass, projectiles, and so forth. If the wheel remains a core technology for humans, perhaps it is for ET as well. Perhaps ET does not possess technologies that are so superior to ours as to allow them to send colonies. For example, we are currently unable to build a fusion reactor, much less be able to fashion one into an interstellar engine, nor do we have any idea how to create enough antimatter for interstellar travel or how to safely segre-gate it from the tanks of ordinary matter within which it must be contained. Just possibly, the engineering challenges are so profound that ET has no practical solutions either. In other words, perhaps there is an asymptote to technological possibilities, and perhaps we are closer to that leveling off than we imagine. Were there no such asymptote to technology, we might observe ET civilizations that had mastered the entire energy output of their galaxies (so called K-III galaxies), but none has been observed [17,18,19]. In fact, no K-II system has been detected either, in which ET controls substantially all of the energy of its home star by enwrapping it within a Dyson sphere [20,21]. The fact that K-II and K-III civilizations and stellar Dyson spheres have not been observed has been taken as evidence that ET does not exist at all. However, this might point to a very different conclusion. First, all such KII-III surveys have been limited, so one can only surmise that KII-III civilizations are not common. However, it may also be that ET civilizations are plentiful but no civilization has technology capable of fully utilizing the enormous energy and material resources of its own star system, much less its whole galaxy, and therefore there are none that feel the resultant pressure to migrate to other stars. Alternatively, perhaps interstellar colonization is all but impossible, and that therefore civilizations moderate their internal growth so as not to exhaust the resources of their solar systems.

### 3.2 Extended Existence Within an Hermetically Sealed World Ship May not Be Possible

We know nothing about ET's own requirements for space travel. For example, whereas humans would require lighting to read by as well as toothbrushes and toothpaste, ET might be bioluminescent and toothless. Nonetheless, by imagining human requirements, and stipulating that ET's may be quite different, we can learn something about the difficulties that would attend interstellar colonization:

**3.2.1** A world ship would have to bring with it a veritable Noah's ark of other life forms. The colonization literature blissfully speaks about interstellar colony percolation and dispersion rates. However, much more than brave astronauts would be required in practice. A proposed colony would need to bring along a viable ecosystem comprised of microbes, plants, insects, fish, mammals, birds, fungi, and so forth. How does one maintain a steady state ecosystem in a hermetically sealed world ship over millenia? How many separate ecosystems can a single world ship maintain? Will one be suitable for penguins and another for parrots? If there will be multiple eco-systems, will there be walls between them as there are none on Earth, such that temperature is maintained at, say, 30° C in an artificial jungle, but at -4° C in an artificial Arctic? On Earth, ecosystems blend together. Will lack of blending present its own set of problems? How many and which species will be brought? No artificial ecosystem could possibly contain all of the richness of its Earth analog. Will future scientists understand ecosystems so well that they will understand precisely which species are critical for its maintenance over millennia? Will they each be suitable for the target planet once it has been terraformed? How might the populations of each species be maintained at zero population growth over the course of the millennia of travel? What will all of these species eat en route and how can their diets be so monitored as to ensure that each species remains intact, e.g., that all the lions do not eat all the sheep? How can 100% efficiency in animal waste and carcass recycling be achieved? Upon arrival at the target planet new problems emerge. A full ark-load of life forms cannot simply be sprinkled about in the hope that they will miraculously fall into a new interlocking ecosystem. In the case of Mars, species might be gradually shipped from Earth over millennia of terraforming (first rudimentary archaea, then bacteria, then algae, then plants, then insects, and so forth up the ladder of life's complexity), but wayfarers to the stars cannot reasonably expect the arrival of wave after wave of subsequent supply ships. Higher species will have to remain on board as the lower species are gradually introduced. The success of the enterprise might depend on the exact ordering of the rollout of species, or worse, the dependencies may be mutual, such that neither an ordering of introductions nor an all at once introduction would work. On Earth, the linkage between species certainly does not flow in a single direction. One might think that elephants should be introduced to the target planet only after their suitable forest habitat has been planted, but some vital plant species may depend on elephant dung for fertilizer, and insects may depend on it for food, while the plants in turn may depend on the insects for pollination. Life is a spaghetti bowl of entanglements such that any proposed ordering of species introductions could result in impossible Catch-22's. The alternative, introducing an ark-load of species all at once, would be virtually assured to lead to the extinction of most of them, while radical speciation is likely for the surviving species, including humans [22]. The problems in transferring an ark-load of life forms from Earth to a planet in another solar system are so legion that the more one ponders the problem the more intractable to solutions the whole enterprise seems to be.

**3.2.2** A world ship mission lasting millennia, including both the travel time and the wait time at destination while the target planet is terraformed, will be subject to internal evolution along multiple parameters [23]. An ecosystem in a hermetically sealed world ship that seems perfectly balanced at the outset might easily fall out of balance, with a flap of the proverbial butterfly's wing. For example, the initial bacterial load might have to be just right down to an exact combination of endless strains. Perhaps the absence of the exactly correct number and types of phages would result in a runaway bacterial imbalance. We know nothing about ET social structure, but human societies cannot possibly remain stable over millennia. Could the population size be maintained within an acceptable range? There may be evolving tribalism and resulting wars. Over millennia, languages mutate so that later generations might not be able to read the operating manuals, much less the literature, of their distant ancestors. Religions and social mores would mutate also. Rules of governance will change. A founding generation's democracy might soon be superseded by dictatorships or theocracies. The founding generation's understanding of and dedication to the mission may not be sustained over many generations. Crucial skills might be irretrievably lost when, for example, all computer scientists are burned at the stake during a generation in which they are accused of witchcraft. Less dramatically, key skills may simply atrophy, such as the way in which blacksmithing has all but vanished today.

**3.2.3** Provisioning such a mega-generational expedition presents potentially insurmountable problems. Consider a single crewmember's quarters. One might imagine a mattress, bed frame, linens, pillows and pajamas. Will these items be made from futuristic materials such that they endure generation after generation? Current bed linens, pillows, and pajamas wear out, so unless we posit radically new materials a source for flax, down and cotton will be required. Does the world ship grow cotton, flax, and raise geese for meat and down feathers? What about a reading lamp; will its metal base rust or fatigue over many generations? Will the light bulb ever need to be replaced?

Entering the bathroom, we are left to wonder whether every generation will use the same toothbrush and whether the water cup will never break. Surely toothpaste is not reusable. What about the medicines in the medicine chest? We must therefore postulate textile mills and innumerable factories to manufacture toothbrushes, toothpaste, light bulbs, lampstands and all manner of pharmaceuticals. A crew member might be expected to use at least 100 toothbrushes in a lifetime. Multiply this by the number of generations. All factories require raw materials. This would either entail that fantastical amounts of raw feeder materials be brought or that virtually every atom in every proverbial toothbrush will be recycled with near perfect efficiency.

**3.2.4** With so many species living in close quarters, the chances of pathogens evolving in one species and then jumping to others is not trivial. There can be horrible knock on effects if one species on which other species depend is wiped out by a pathogen.

**3.2.5** It is hard to imagine how genetic variability could be maintained across each and every species that resides in the world ship ark.

**3.2.6** All the generations of humans and all the generations of animals will have to be fed. That will mean that vast tracts of land must be dedicated to agriculture. That will in turn require interior illumination as bright as our Sun. When considering the amount of fuel necessary for a journey of millenia, one must take into account not only the amount necessary to accelerate and decelerate the ship to and from cruising speed, but also the amount of energy required for all manner of internal use, chief among which will probably be for agriculture.

**3.2.7** The world ship literature has ignored the possibility that the size of the proposed ark might be radically reduced by sending frozen bacteria, plant seeds, mammalian zygotes and so forth, instead of mature living beings. In fact, ET might simply send DNA sequences in the form of computer code, along with a gene sequencing machine to translate the code into viable life forms upon arrival at the target planet. However, this may pose its own insoluble problems. For example, a zygote is far more than a DNA sequence. It is a functioning cell with a complex cell membrane and cellular machinery additional to its DNA. It is not assured that a human zygote can be brought to full term in an artificial womb, lacking a very particular maternal chemical, hormonal and bacterial environment. Myriad and highly specialized wombs would have to be transported for each proposed species of mammal. Mammals require very special handling that would be difficult for any robot to emulate, including the warmth, smell, softness and sounds of a mother. In fact, human babies cannot live at all without constant care. Consequently, humanlike robots would seem to be a necessity. There would also have to be robots that simulate virtually every higher mammalian species (elephants require mothers also). If all of the species are to be brought to life within the world ship, then that ship will have to be about the same size as what has been previously imagined in order to accommodate them. Those first babies would have to be born into a fully functioning and mature world with, for example, robotically run agriculture and dairy farms to feed them, and robotic parents to discipline and school them. Alternatively, if they are to be brought to life only upon their introduction on the target world, all the issues surrounding species introductions would still obtain and all issues around terraforming would have to be dealt with solely by AI robots.

**3.2.8** The number of other conceivable world ship failure points staggers the mind: micro-meteor impacts, gamma ray burst, high energy cosmic rays, life system support breakdowns, computer viruses, and social dislocations of every conceivable type. In sum, there are so many potential astronomical, technological, life system, and sociological concerns for such a mission that in their accumulation over millennia (it may only require a single failure point to ruin a mission) the net result may be that there would be virtually no chance for a world ship to successfully arrive at its destination.

### 3.3   Sorry, But This Planet is Taken

However, if the above arguments are surmountable, and world ships are indeed viable, then, as Hart and others have suggested, perhaps a fast ET colonizing world ship could arrive at a nearby star in a reasonable amount of time. However, where would it land? If there is a high probability of life endogenously arising on rocky planets located within a star's habitable zone (HZ), then suitable targets for colonization might already be inhabited. If that local life is compatible with ET's, its microbes might eat or infect the colonists. The chances that ET's life is fundamentally identical to that of the target planet would be remote. Even if both were based upon carbon and DNA, the colonizing ET and the local life might be built upon a different foundation of amino acids. Life on Earth utilizes just 22 of the more than 500 known amino acids. Would be ET colonists and the local life forms might utilize different sets of amino acids, or one or the other or both might not utilize amino acids at all. Similarly, the two life forms might utilize very different collections of key proteins, have opposite chirality, and so forth. However, it might still be possible that local microbes could eat the colonists by breaking its chemistry down into root component parts such as simple sugars [24]. H.G. Wells' *War of the Worlds* paradigm may be all too real. On the other hand, if the two life forms are not at all compatible such that one cannot ingest the other, the colonists would likely starve, because the local life would already have filled all viable ecological niches. For example, ET's equivalent of lettuce and tomatoes might not grow in soil infused with alien microbes. A life-bearing world, be that life as we know it or not, will almost certainly have an atmosphere that is much different from that of the colonists, if only in its atmospheric pressure and relative percentages of gases such as oxygen, nitrogen, carbon dioxide, methane and so forth. ET would almost certainly not be able to breathe it. If it were to alter the atmosphere to its specifications, it would presumably poison the indigenous population. Is that what it would want to do? Would galactic metalaw constrain ET colonists from attempting this?

### 3.4   Terraforming Might Solve Some Problems But Not Others

If life is rare, the target planet may be barren, and therefore, possibly in accord with galactic metalaw, free for colonization. In that case, colonists will be able to commence terraformation.

**3.4.1.** Terraformation could take at least many thousands of years, using any imaginable technology. The requirements for terraforming Mars have been well considered in the literature, and are instructive. First, massive amounts of $CO_2$ would have to be released from the regolith and polar caps in order to create a necessary warming greenhouse effect and sufficient atmospheric pressure for humans to be able to exist on Mars' surface. Zubrin and McKay [25], terraforming optimists, suggest that the best way to do this would be to position a 100 km radii mirror weighing 200,000 tonnes in Mars orbit in order to melt the polar dry ice. However, known $CO_2$ reservoirs in the ice caps and the regolith taken together are insufficient to raise atmospheric pressure to more than a small fraction of the levels necessary for human physiology. Consequently, very deep mining for possible limestone deposits of $CO_2$ would be required. Massive amounts of energy would be required to heat the rock, releasing its gases. However, this effort is merely in the cause of raising atmospheric pressure to human/animal tolerances. Because such a high concentration of $CO_2$ is poisonous for animals and most plants, oxygen tanks would be required not just for human adults, but for babies and elephants and everything else. When considering a planet in another solar system, such an orbital mirror and vast quantities of excavating equipment would either have to be included in the inventory of the world ship, or built upon arrival. Continuing the story, once atmospheric pressure has been raised, most of the $CO_2$ would still need to be converted to $O_2$. The only known means to produce atmospheric quantities of oxygen from $CO_2$ relies upon biological pathways. McKay has estimated that it would likely take at least hundreds of thousands of years for primitive plants and algae to accomplish this [26]. Were this a similar situation to the one interstellar colonists would encounter, all the while the world ship's menagerie of plants and animals would have to wait on board or in bubbled enclosures on the surface. If a world ship must wait in orbit for many thousands of years, or more, then why should it not simply persist as a world ship, and forgo colonization altogether? After all, the simulated mixture of gases in the world ship (for humans, 78% nitrogen, 21% oxygen, 1% water vapor, etc.), its atmospheric pressure, its light/dark cycle (for humans, 24 hours), as well as its artificial centrifugally created gravity (for humans, g=1) would be ideally set to mimic home.

**3.4.2** Terraforming cannot solve for an undesirable amount of gravity. Could ET colonize a planet with twice or half its home planet's gravity? It seems obvious that it would be difficult, if not impossible, for humans to live on a planet with twice Earth's gravity. An average-sized human would weigh some 350 pounds, and lack compensating musculature. Would a sub-Earth be any better (Mars' gravity is 38% of Earth's)? We simply do not know anything about the effects of a much-reduced gravity over a human lifetime. We know some things about short duration microgravity effects, for example, from sojourns on the International Space Station, and they are not good. What of the effects on the rest of the life forms in the world ship ark? Would spiders weave webs on a sub-Earth like Mars? Would elephants develop a new bouncing gait? Would birds fly ten times higher or crash and break their necks? Who knows? Terraforming Mars or similarly sized sub-Earths in other solar systems would presumably require an enhanced atmosphere (see above). However, if gravity was too weak to resist the solar wind and hold on to Mars' original atmosphere, how will it, or a similar sub-Earth, hold on to an artificially created atmosphere?

**3.4.3** Terraforming cannot solve for an inadequate surface chemistry. Life on Earth requires a host of elements, such as boron, nitrogen, iron, iodine, phosphorous, zinc, carbon, calcium, and potassium; and of course, there must an adequate quantity of water. If even one of these elements or water is missing, Earth life might not be able to establish itself on the target planet.

**3.4.4** Colonists may have to take into consideration the chronobiological effects of a light/dark cycle much different than 24 hours. By happy coincidence, the planet most often considered for colonization and terraforming in our Solar System is Mars, which at 24 hours 40 minutes just happens to have a day/night cycle that is very close to Earth's. However, the nearest Earth analog outside of our Solar System may have a 7-hour day or a 50-hour day, with huge deleterious effects on human biology, not to mention on the world ship's ark full of other life forms.

**3.4.5** Whereas there might be viable and practical solutions to some problems associated with terraforming, it is hard to imagine technical solutions to all problems. Indeed, were ET determined to terraform another planet to its specifications, despite all of the above, then to do so would be vastly easier to accomplish within one's own home solar system, where it could presumably abide the thousands to millions of years required to accomplish the task on its home planet.

### 3.5 It May Be Naïve to Think That Nearby Stars Contain Truly Habitable Planets

The distance to the nearest genuinely viable planet (i.e., right gravity, right warmth, right day/night cycle, right atmospheric composition and pressure, right axial tilt, right eccentricity, right surface chemistry, tectonic plates enabling an efficient recycling of $CO_2$, uninhabited, not a waterworld, etc.) may be impractically far away for travel ("impractical" in the sense that, although not strictly impossible in accordance with the laws of physics, it is all but impossible nonetheless because of constraints upon even advanced technologies). There are estimated to be 50 billion rocky planets within the HZs of their stars in the Milky Way [27], leading Marcy to estimate that approximately 11% of G-stars have rocky planets smaller than 1.5 Earth radii orbiting within their HZs [28]. This would imply that most alien civilizations would not need to travel more than 10-15 light years in order to find one. However, when one further constrains the definition of "habitable" to also mean "habitable for ET's biological colonists," namely, that it has a gravity, temperature, atmospheric composition and pressure, surface chemistry, and length of day that is compatible with ET's home planet, and on top of all this that it also uninhabited, then the distance to the nearest candidate may be impractically far.

### 3.6 ET Cannot or Does Not Take on the Attributes of a Virus

Intelligent life need not be virus-like, simply multiplying *ad nauseum*. Human birthrates have dropped precipitously with wealth and education. The size of computers has declined just as their computing power has exponentially increased. It is possible, as Hart [29] and others [30] have argued, that it might take less than a mere 50 million years to colonize the entire galaxy, but what then? Would a virus-like ET burn itself out? Exponential population growth on planet Earth would be unsustainable. Surely, humanity would be doomed were such growth not held in check. The same principle applies to the galaxy as a whole. Endlessly replicating von Neumann machines, multiplying like the Micky's brooms in the "Sorcerer's Apprentice," would eventually consume all matter useful to it in the galaxy. Obviously, this has not happened in the Milky Way, and no evidence exists for gobbled up galaxies elsewhere in the universe. It may be that any intelligent species that could theoretically manage to colonize the galaxy has already grappled with and eliminated any tendency toward exponential growth [31]. Humanity may already be on a path to sustainability—or not. There are trends in both directions, with our very fate held in the balance. This question should be settled either by our self-destruction or not, long before we have any practical ability to send colonists to the stars. I have previously argued [8] that galactic metalaw may consist of a few words only, roughly translated into English as "don't come;" "stay still;" or "you move—you die." The galactic com-munications network comprised of probes tied together by communication nodes may not only serve communication, but also to enforce rules meant to guard the galaxy against any runaway virus-like infestation. The colonization of even a single other star system might be *prima facia* evidence that the civilization is virus-like in nature, warranting its immediate destruction. If carbon-based life uploads itself into silicon AI based life, then there would be no apparent need to colonize for the sake of *lebensraum*. There would be no Darwinian selection pressures, no maturation and old age, and no need for the AI beings to sexually reproduce or to mutate [32]. The AI beings might simply add ET's version of CPU and disk space large enough to do the calculations that give them some type of internal meaning or pleasure. Colonization would serve no apparent purpose.

## 4 IF ET IS NEITHER HERE NOR EN ROUTE, WHERE IS IT?

ET may have ample sunlight and raw materials in its own solar system.

**4.1** Drake made the argument early on that ET has not come to our Solar System because it is sunning itself contentedly in its own solar system [33]. His logic seems correct. The Earth receives only about one one-billionth of the Sun's total luminosity. In the event that one one-billionth of the Sun's energy becomes insufficient for Earth's needs, humans can always harvest more energy, ultimately encircling the Sun with a Dyson sphere. Only when a Dyson sphere, harvesting substantially all of its star's energy, proves insufficient would a civilization feel compelled to move on to another star system.

**4.2** ET may also not be motivated to leave its home star system for lack of material resources. For example, in our Solar System, the metallic asteroid, Psyche, is estimated to possess at least one million times the metals that have ever been mined on Earth [34]. If ET's star system is also the site of such magnificent abundance, there would seem to be no impetus to leave for want of it.

**4.3** Even if ET feels impelled to leave its home planet and travel for vast amounts of time in a world ship in order to arrive at a planet, and then spend additional vast amounts of time in the effort to terraform that planet, it may be that ET is well suited to live in a multigenerational world ship. ET might simply continue to live in interstellar space, or directly orbit a star to harvest its energy without need to colonize any of its planets [35].

## 5 BUT THE END OF THE WORLD IS WELL NIGH!

There is a reasonable objection to all of the preceding lines of reasoning: G-stars are doomed to die. During its red giant (RG) phase, a life-bearing planet in the traditional HZ (a planet within a radius from its star allowing for liquid water on its surface) would be sterilized. Superficially, this would seem to demand that any ET civilization based on a planet located within the HZ of a G-star would be forced to migrate to a new star system [36]. However, there are alternative and possibly better solutions:

### 5.1 Migration Within the Planetary System and Astro-engineering

As its home world heats up, ET might migrate to its outer solar system, either residing in world ships or settling on planets, moons, or asteroids within the new HZ, which would be reset at approximately 7–15 AU for a RG spawned by a G-star. Given billions of years of foreknowledge, it may even be feasible to gradually increase the radius of ET's home planet's orbit toward a new HZ defined by the RG, and then shrink the orbit again as the star enters the WD stage [37,38]. A G-star enters its RG stage once it has exhausted a mere 10% of its supply of hydrogen during the main sequence phase. The balance remains outside of the core, where pressures are insufficient for nuclear fusion. An advanced ET might devise a means of shunting more hydrogen from the star's outer layers into the core and thus greatly prolong its main sequence life [39]. ET might also cool its star by reducing its opacity [40]. Alternatively, as the star sloughs its outer layers and enters into its WD stage, ET could remain in world ships or on planets, moons or asteroids, that is, basically anywhere in the system. It might even reinhabit its world of origin, harvesting and beaming energy from the WD back to it. Of course, all other viable energy sources might be used, such as nuclear energy. However, no matter where the remnant ET civilization orbits around a WD, it would need to be shielded from intense UV. It could, though, convert those high energy photons into useful energy.

Any civilization that survives the RG phase and persists into the WD stage is presumably an ancient and advanced civilization. M-stars have been postulated to be a good SETI target class. Because of their longevity, older ones might also be home to ancient civilizations, and these civilizations would not yet have suffered the stress of the departure of their star from the main sequence. However, we only know with certainty that life can arise within a G-star system. M-stars pose difficulties, as it is unclear how life could evolve in the presence of their intense stellar flares. Furthermore, the HZ around an M-star is precariously thin.

### 5.2 In No Event Would Hart Be Vindicated

Discounting all of the arguments in this paper, even if it is deemed necessary to migrate away from one's own star system and completely feasible to colonize distal star systems, simply relocating to the next appropriate star system would suffice, with no apparent need to exponentially colonize the entire galaxy. Such a civilization might star hop at the rate of once in every several billion years, far less than the rate imagined by Hart and his school of thought.

## 6 SEARCH STRATEGIES

### 6.1 Targeting Red Giants and White Dwarfs

The two main conclusions of this paper, namely, (a) that colonization, or at least the transport of all biological beings to another star system, may be virtually impossible; and (b) that survival within a dying star system may be feasible, suggests new search strategies. If there is no viable exit plan, then ET must make the best of a bad situation and find a way to persist in proximity to RGs and WDs. Almost by definition, such persisting civilizations should be among the oldest, and perhaps therefore self-confident, yet they would also undeniably be experiencing the distress of their dire straits. As a consequence of both their self-confidence and/or their dire situation, they may be the most likely to attempt to communicate with beacons transmitted from within their own star systems.

Approximately 5% of stars in the Milky Way are G-stars. However, approximately 15% are WDs, ranging from ~ .17 – 1.33 $M_\odot$, with a median mass of ~ .6 $M_\odot$. This rich target source has been all but ignored in previous SETI searches, though BreakThrough Listen has recently added some WDs to its target lists [26]. Therefore, it is recommended that the typical SETI target list that prioritizes G-stars, with some K and M stars and perhaps a small smattering of WDs, be reverse ordered to prioritize RGs and WDs above all other targets, followed by the oldest G and K stars. Since there is a direct correlation between the size of the progenitor star and the mass of the WD, this new target list might be further refined to focus primarily on WDs whose progenitor stars were approximately the size of the Sun. WDs whose progenitors were G2 stars like the Sun would have been formed about 10 billion years ago. Because the very early universe was dominated by supermassive stars and their attendant supernovae blasts, metallicity was likely to have been sufficient 8 - 11 billion years ago to allow for the formation of rocky planets. Indeed, rocky debris has been observed in the spectra of WDs, presumably the remnants of shredded asteroids and rocky planets such as Earth and Mars [41]. Nonetheless, the preponderance of WDs derive from progenitor stars larger than the Sun that lay within the 1–1.3 $M_{\odot}$ range. This does not itself greatly impact the thesis, as ET might have arisen on a planet in the HZ of a star in this range. Although G-stars more massive than the Sun have somewhat shorter lifetimes on the main sequence, we already know that our technological civilization arose in only 4.5 billion years. These stars satisfy that condition.

Even if interstellar migration is feasible, it would presumably fall far short of rescuing the entire native biome. In terms of Earth, we can imagine sending thousands or tens of thousands of colonists to Alpha Centauri, or maybe even some millions of colonists in great fleets of word ships, but not the entire human population of Earth. What fate would await the billions of people who would inevitably be left behind? Are they to submissively perish, or attempt, as argued here, to find survival solutions within our own Solar System? And what about the rest of the biome, colloquially, the lions and tigers and bears, as well as the birds and the bees and the flowers and the trees? Clearly, an *in situ* rescue plan is the only viable rescue plan. This line of argument also leads to the conclusion that the best SETI targets may be end stage star systems. A focus on end-stage star systems suggests some search strategies:

**6.1.1** During the WD phase onset, ET might relocate from its outer solar system abode to its planet of origin. That planet's surface would have certainly been sterilized during the RG phase, and frozen during the WD phase. It is an open question as to whether it would also have been stripped of its rocky mantle [41]. Provided that the planet has survived intact, it will have two clear advantages, namely, that its gravity and its day/night cycle would be exactly right for ET (the WD flux would have to be lensed to recreate the day/night cycle). Deep ground water might also persist. Its atmosphere would have to be restored, but this would be vastly easier to accomplish than traveling to a nearby star system to do essentially the same thing. Energy could be harvested from the WD and transmitted via microwaves or lasers. As a search strategy, one might look for spillover from these beams of energy, transits of massive power stations, or WDs that are anomalous in their energy output, for example, dimmer than models indicate.

**6.1.2** Young WDs have surface temperatures of 50,000 to 100,000 K that emit deadly amounts of UV. Therefore, strictly speaking, there can be no HZ around a WD in the absence of some sort of artificial UV protection. Perhaps giant UV filters might be detectable, though they would likely orbit the planet rather than the star. The UV problem also militates in favor of ET remaining in the outer Solar System or at least approaching no closer to the WD than its planet of origin, which would have migrated outwards to approximately twice the distance of its previous orbit because of the WD's weaker gravity relative to its progenitor star.

**6.1.3** If adequate UV filtering can be achieved, ET might establish itself in the WD's new HZ of ~.03 AU. Such "planets" orbiting improbably close to a WD might in fact be artificial world ships, or they could be actual planets artificially steered into the WD's HZ. 1.3% of such objects in the HZ, assuming randomly oriented orbits, would transit Earth's line of sight [42]. Transits would last about 50 seconds. An Earth-sized object orbiting an Earth-sized WD would cause an easily detectable 50% dimming. In fact, objects as small as 200 km. would have been detectable by Kepler, though none were observed. Consequently, a program to detect objects in the HZ of WDs might be undertaken. On the one hand, lacking Kepler's wide field of view, such a program would presumably have to target stars individually. On the other hand, an object at .03 AU completes one orbit in only 2.7 days. Three transits are required to describe an orbiting object. It would require about ten years of telescope time to complete a survey for transits of the nearest 1000 WDs. This could be accomplished by .3 m telescopes, though because they are Earth-bound rather than space-based like Kepler, multiple telescopes spaced around the world would be required to be tethered together to accomplish this [43]. In the event that objects were discovered, one telltale sign of artificiality would be a lack of tidal locking, though tidal locking cannot be readily determined by the current generation of telescopes.

**6.1.4** Search for WDs with anomalous spectra, such as would be the case if they were laced with the waste products of nuclear fission or fusion from power plants, such as radioactive nuclides 232Th, 235U, or 238U or had non-astrophysical emission lines [28].

**6.1.5** Older G-stars bear special scrutiny for another reason. A normal G-star burns only about 10% of its hydrogen before leaving the main sequence and becoming an RG. This is because 90% of the hydrogen resides in its outer layers, rather than in the star's core where temperatures and pressures are high enough to fuse hydrogen into helium. A G-star could be made to increase its lifetime on the main sequence were hydrogen artificially funneled from the outer layers of the star into its core. However, the required engineering would probably rely on new physics, and it is not clear how a G-star that had been artificially altered might detectably differ from a normal G-star.

**6.1.6** Maoz, et al. [44] observed subtle sinusoidal luminosity periodicities in 7 out of 14 observed WDs. Only tentative explanations were proffered, and then no one explanation fit more than one or two WDs. The authors did not consider artificial explanations for the observed periodic dimming, such as if they were caused by swarms of world ships or ET power stations. Further data on WD periodic dimming should be gathered and explanatory models refined with an eye toward nullifying a hypothesis of artificiality.

**6.1.7** Although no planet has yet been observed in orbit around a WD, an apparent debris disc has been observed in a ~4.5 hour orbit around WD 1145+017 located in the Kepler 2 field [45]. This has generally been interpreted as resulting from an asteroid or rocky planet that, after having migrated close to the WD, is currently in process of being shredded by tidal forces.

However, no one has tested the possibility that this apparent debris cloud could indicate astro-engineering in progress, such as the construction of world ships, power plants or a Dyson sphere. The two interpretations are not mutually exclusive, in that an advanced civilization might intentionally steer a rocky body toward a WD so that it might be tidally shredded and its component materials thereby more easily utilized for ET's purposes. ET would then mine the useful materials while allowing the rest to orbitally decay and infall onto the WD.

**6.1.8** Because WD pollution has been well established, the question arises as to where the pollution comes from. The standard, though as yet untested, assumption is that rocky body orbital interactions in a system that has undergone gravitational upset as it transitions from G ➔ RG ➔ WD cause asteroids or other rocky bodies to infall onto the WD, polluting its surface. Computer modelling can determine whether the observed WD metallicity can be explained solely by orbital dynamics. However, this is a task for future astronomers. The only solar system that can be modeled at the moment is our own, and it appears atypical. In order to model other solar systems we would have to possess a complete inventory of their planets and asteroid belts (if any). Currently, we know very little about planets further out than ~1AU nor about exo-rocky bodies smaller than the Earth. If we were able to fully describe a large sample of G-stellar systems, run computer models, and find that the amount of debris observed on actual WDs is larger than the amount of debris predicted by models, perhaps artificiality could explain the difference. Until such future time as enough stellar systems are fully described tentative models might be constructed from our own Solar System, plus hypothetical solar systems constructed for the purpose. There is an important caveat. Although in the future we may be able to model the fate of rocky objects that are actually observed in G-star systems, the results can only cautiously be applied to current WD systems, as these systems may have begun their life with less metallicity than current G-stars, skewing results.

**6.1.9** WD pollution and WD dimming might be examined for spatial or temporal patterns. It is conceivable that an ET might deflect debris intentionally into its WD, or modulate its dimming, as a way to signal for attention, perhaps an SOS [28]. As opposed to directed EM beacons, such messages would be observable from most directions.

**6.1.10** Planetary nebula, the intermediary phase between RG and WD, may be of particular interest. There could be a frenzy of detectable ET astro-engineering marking the transition between RG and WD stages.

**6.1.11** The above analysis also applies to late stage G-stars. G-star flux increases over time. Up until the present, there has been a Gaia-like thermostatic feedback loop that has kept Earth's temperature within a range that has allowed for life to persist either on its surface, or, in the extreme circumstances of a snowball Earth, at least in the liquid oceans beneath the ice. However, the Sun will double in luminosity over the next one billion years. As a result, the Gaia effect will likely fail Earth within about the next 500 million years, by which time the Sun will have caused Earth's oceans to boil away. Such a fate might be avoided if huge space mirrors deflect a sufficient amount of sunlight away from Earth, or the Earth itself is intentionally maneuvered further away from the Sun [26]. Such mirrors might be detectable by future generations of telescopes, while subtle outward migrations of exoplanet orbits (to maintain its presence within an expanding HZ) might also be detectable.

### 6.2 The Possible Implication of the Discovery of Earth-like Life on Enceladus, Europa, Titan, Pluto and Other Outer Solar System Objects

If colonization by even an advanced ET species is not possible, while at the same time persistence within one's own solar system also cannot be achieved past the RG and WD stages, then ET might be forced to concede that it is doomed as a species. However, ET might at least be able to facilitate the long-term persistence of life. ET might send probes filled with an assortment of microbes to populate newly born planets, letting them evolve in their new homes as they might. Even if advanced life could not possibly colonize an analog of the very early Earth, microbes might. Consequently, we may be ET, the product of directed panspermia more than 3.7 billion years ago by an earlier and now extinct civilization. Therefore, if Earth-like life (i.e., using the same set of amino acids, based upon DNA and/or RNA, and so forth) is found on such outer Solar System objects as Enceladus, Europa, Titan or Pluto, it would greatly strengthen the inferential case for interstellar probes, though it would not foreclose other explanations, such as non-directed interstellar or intrastellar panspermia, as being the correct explanation.

## 7 CONCLUSIONS

Faced with the inevitable phase transition of its parent star from the main sequence to RG and from RG to WD, ET would be faced with three stark choices: (a) passively succumb to extinction; (b) colonize a nearby star system (or simply cruise indefinitely in interstellar space); or (c) shelter in place. It is the position of this paper that taking the necessary measures needed to continue to exist within one's home system is by far easier than colonizing, and presumably preferable to passive extinction. ET world ships have not been observed in our Solar System [9], nor is there any evidence of past colonization. It is insufficient to suggest that colonization is not possible due to either the immensity in the amount of energy it would require at relativistic speeds to make the journey, or because of the immense amount of time a journey would take at nonrelativistic speeds, such as the velocities achieved by our own interstellar space probes, Pioneer, Voyager, and New Horizons. Interstellar journeys would demand inordinate amounts of time, energy, or both, but they may not be impossible on those grounds alone. However, even were ET willing and able to spend the time and/or energy necessary for interstellar travel, it would still be stymied by some or all of the other factors enumerated in this paper, which must be regarded cumulatively.

If ET has the technology to create world ships in which it can survive for at least the tens of thousands of years it would take to both travel to another star system and then terraform an appropriate planet therein, then it begs the question of why would they bother landing on a planet at the end of this time, rather than simply persist in a world ship where the gravity, atmosphere, day/night cycle and so forth have been idealized? If there is no particularly good reason to disembark on a planet in a distal star system, then why bother to travel there in the first place? Certain solutions to practical problems might themselves not violate any law of physics, but are so economically preposterous that they are never put into practice. For instance, no one flies to the other side of the Earth to buy a loaf of bread. Rube Goldberg made a career of inventing fantastically complex and silly solutions to simple problems; ET would be smarter than that.

SETI searches should reasonably target G-stars preferential-ly in accordance with an increase in their ages. For example, there would be little apparent reason to target stars of ages < 1 billion years, since those stars would be unlikely to harbor native technological civilizations. If migration is not an option, as argued in this paper, then G-stars should be targeted in increasingly larger bins as they age, with the greatest preference given to the oldest G-stars. There is no reason to end this sliding scale with the demise of the G-star. It should be continued through the subgiant, RG, planetary nebula and WD stages.

This same argument holds true even if one disagrees with the conclusion that interstellar colonization is virtually impossible. One can imagine a single world ship, two, or a dozen making the one-way voyage to another star, but not world ships containing the entire native technological population, along with the entirety of its biomass. Would those who remain behind resign themselves to annihilation or dodge the bullet by migrating away from their star and toward their outer solar system during the RG phase, and then inwards again at the onset of the WD stage?

The arguments proffered here apply only to biological ET colonists. The AI probe hypothesis posits that migratory visits by ET probes are quite feasible, as probes would be impervious to the needs of biological creatures for a correct gravity, atmosphere, day/night cycle, flight duration, life support systems, and so forth. Unlike biological beings, they would also be impervious to the passage of deep time. Without a requirement for complex life support systems, a probe would be orders of magnitude less massive than a world ship.

Such beings may require some raw materials for self-repair or build out of capabilities, and might therefore land on asteroids. But they may have no need whatsoever to dive into a deep gravity well such as Earth. Not being subject to Darwinian evolution, it is hard to imagine why they would be programmed to reproduce endlessly, and, even if von Neumann replicators are possible, they may be outlawed in accordance with galactic metalaw.

### Acknowledgement

The author gratefully acknowledges the kind assistance of Geoff Marcy in the organization of the ideas expressed in this paper.